\documentclass[]{spie}  %>>> use for US letter paper
\usepackage{amsmath,amsfonts,amssymb}
\usepackage{graphicx, array}
\usepackage{physics}
\usepackage{float}
\usepackage{pgfplots}
\usetikzlibrary{decorations.pathreplacing}
\usepackage[colorlinks=true, allcolors=blue]{hyperref}

\title{Evaluating Transformer-based Semantic Segmentation Networks for Pathological Image Segmentation}

\author[a]{Cam Nguyen }
\author[a]{Zuhayr Asad}
\author[a]{Yuankai Huo}
\affil[a]{Department of Computer Science, Vanderbilt University, Nashville, TN, 37235 USA}

\authorinfo{Corresponding author: Yuankai Huo: E-mail: yuankai.huo@vanderbilt.edu}

\begin{document} 
\maketitle

\begin{abstract}

Histopathology has played an essential role in cancer diagnosis. With the rapid advances in convolutional neural networks (CNN). Various CNN-based automated pathological image segmentation approaches have been developed in computer-assisted pathological image analysis. In the past few years, Transformer neural networks (Transformer) have shown the unique merit of capturing the global long-distance dependencies across the entire image as a new deep learning paradigm. Such merit is appealing for exploring spatially heterogeneous pathological images. However, there have been very few, if any, studies that have systematically evaluated the current Transformer-based approaches in pathological image segmentation. To assess the performance of Transformer segmentation models on whole slide images (WSI), we quantitatively evaluated six prevalent transformer-based models on tumor segmentation, using the widely used PAIP liver histopathological dataset. For a more comprehensive analysis, we also compare the transformer-based models with six major traditional CNN-based models. The results show that the Transformer-based models exhibit a general superior performance over the CNN-based models. In particular, Segmenter, Swin-Transformer and TransUNet-all transformer-based-came out as the best performers among the twelve evaluated models. \\

\end{abstract}

% Include a list of keywords after the abstract 

\section{INTRODUCTION}
\label{sec:intro}  % \label{} allows reference to this section

Digital pathological image analysis where diseased tissues are examined under WSI leads to a paradigm shift in clinical practice that allows the pathologists to assess the digitized tissues remotely. Moreover, computer-aided methods can be deployed on the digital WSI to further assist and practically automate the quantification of cancer tissues for histopathologists, which is currently a tedious and resource-intensive process\cite{hel}.The past few years had witnessed a tremendous success of deep learning in digital pathology. In particular, researchers have developed and deployed a convolutional neural network (CNN) with the encoder-decoder architecture to segment tumor tissues. Notable approaches include U-Net\cite{unet}, DeepLabV3\cite{DeepLab}, PAN\cite{pan}, etc. Limited by the nature of the local convolution operations, CNNs tend to focus on local details in segmentation, without capturing the global information\cite{huo2021ai,liu2021simtriplet}. Recently, Transformers have emerged as a new deep learning paradigm for semantic segmentation. Moreover, recent studies have shown that the Transformers can achieve superior performance as compared with the CNN-based approaches in various semantic segmentation applications\cite{vit, beit, sunet, tunet, medt, segm, swin-trans}. The major advantage of Transformers over CNN is their superior ability in capturing the global contextual information of an image, including long-distance relationships and dependencies. The state-of-the-art Transformer-based semantic segmentation methods can be roughly ascribed to two categories: (1) convolution-free design \cite{segm, sunet, swin-trans} and (2) CNN-Transformer hybrid design \cite{tunet, medt}. Besides this, the Transformer can be used as a pre-training tool to learn a representation model for the downstream semantic segmentation\cite{beit}. 

\noindent Even with promising results in a variety of semantic segmentation tasks, there are very few, if any, existing studies that systematically evaluate the Transformers for histopathological image segmentation. In this paper, we provide a quantitative performance evaluation of six recent transformer models and compare them with six traditional CNN-based methods (total of 12 segmentation methods)-for tumor segmentation using the liver histopathological images (Table.\ref{tab:my_label}). 

% The paper is organized as follows. Section \ref{sec:Method} describes the uniformed analysis pipeline, from prepossessing to evaluation, for evaluating 12 semantic segmentation networks. Section \ref{sec:Experiments} describes an experiment in which our pipeline is used to evaluate different semantic segmentation models. Section \ref{sec:Results} presents the results. \\

% \begin{figure}[H]
% \begin{center}

% \includegraphics[height=3.5cm]{fig/0117img.pdf}
% \includegraphics[height=3.5cm]{fig/0117gt.pdf}
% \includegraphics[height=3.5cm]{fig/0117pred.pdf}

% \end{center}

% %{ \label{fig:1stresultimage} 
% %This upper panel shows an example of pathological tumor segmentation in a whole slide image (WSI) from the PAIP liver tumor dataset~\cite{paip}, using Segmenter~\cite{segm}. The lower panel lists the six CNN methods and six Transformer methods that are evaluated in this study.}
% \end{figure} 

\begin{table}[t]\small
    \centering
    \begin{tabular}{p{4.5cm}p{1cm}p{0.5cm}p{9.3cm}}
    \hline
    Transformer-Based Models     & Acronym   & Year   & Download URL \\
    \hline
    TransUNet\cite{tunet}  & TUNet & 2021  & \url{https://github.com/Beckschen/TransUNet} \\ 
    Swin-Unet\cite{sunet} & SUnet & 2021 & \url{https://github.com/HuCaoFighting/Swin-Unet}\\
    Swin-Transformer\cite{swin-trans} & Swin-Trans & 2021 & \url{https://github.com/SwinTransformer/Swin-Transformer-Object-Detection}\\
    Segmenter\cite{segm} & Segm & 2021 & \url{https://github.com/rstrudel/segmenter}\\
    Medical-Transformer\cite{medt} & MedT &  2021 & \url{https://github.com/jeya-maria-jose/Medical-Transformer} \\
    BERT Image Transformers\cite{beit}  & BEiT & 2021 & \url{https://github.com/microsoft/unilm/tree/master/beit}\\
    \hline
    CNN-Based Models  & & & \\ \hline
    Pyramid Scene Parsing Net\cite{psp} & PSPNet & 2016 & \url{https://pypi.org/project/segmentation-models-pytorch/}\\
    U-Net\cite{unet} & U-Net & 2015 & \url{https://pypi.org/project/segmentation-models-pytorch/} \\
    DeepLabV3\cite{DeepLab}  & DLV3 & 2016 & \url{https://pypi.org/project/segmentation-models-pytorch/} \\
    Feature Pyramid Network\cite{fpn} & FPN & 2016 & \url{https://pypi.org/project/segmentation-models-pytorch/} \\
    Pyramid Attention Network\cite{pan} & PAN & 2018 & \url{https://pypi.org/project/segmentation-models-pytorch/}\\
    LinkNet\cite{linknet} & LinkNet & 2018 & \url{https://pypi.org/project/segmentation-models-pytorch/}\\ 
    \hline
    \end{tabular}
    \caption{This table lists the six CNN methods and six Transformer methods that are evaluated in this study.}
    \label{tab:my_label}
\end{table}

\section{Method}
\label{sec:Method}

% \subsection{Tumor Segmentation Pipeline}
% \label{sec:Cancer Segmentation Pipeline}

% The same tumor segmentation pipeline is used to evaluate different semantic segmentation networks (SSN). As illustrated in  the input of the algorithm is an a WSI. Then, the WSI is tiled into smaller patches at either 20$\times$ or 40$\times$ with the same patch size 512$\times$512. 

\subsection{Overall Framework}
\label{sec:Overall Framework}

\begin{figure}[H]
    \centering
    \includegraphics[width=6.5in]{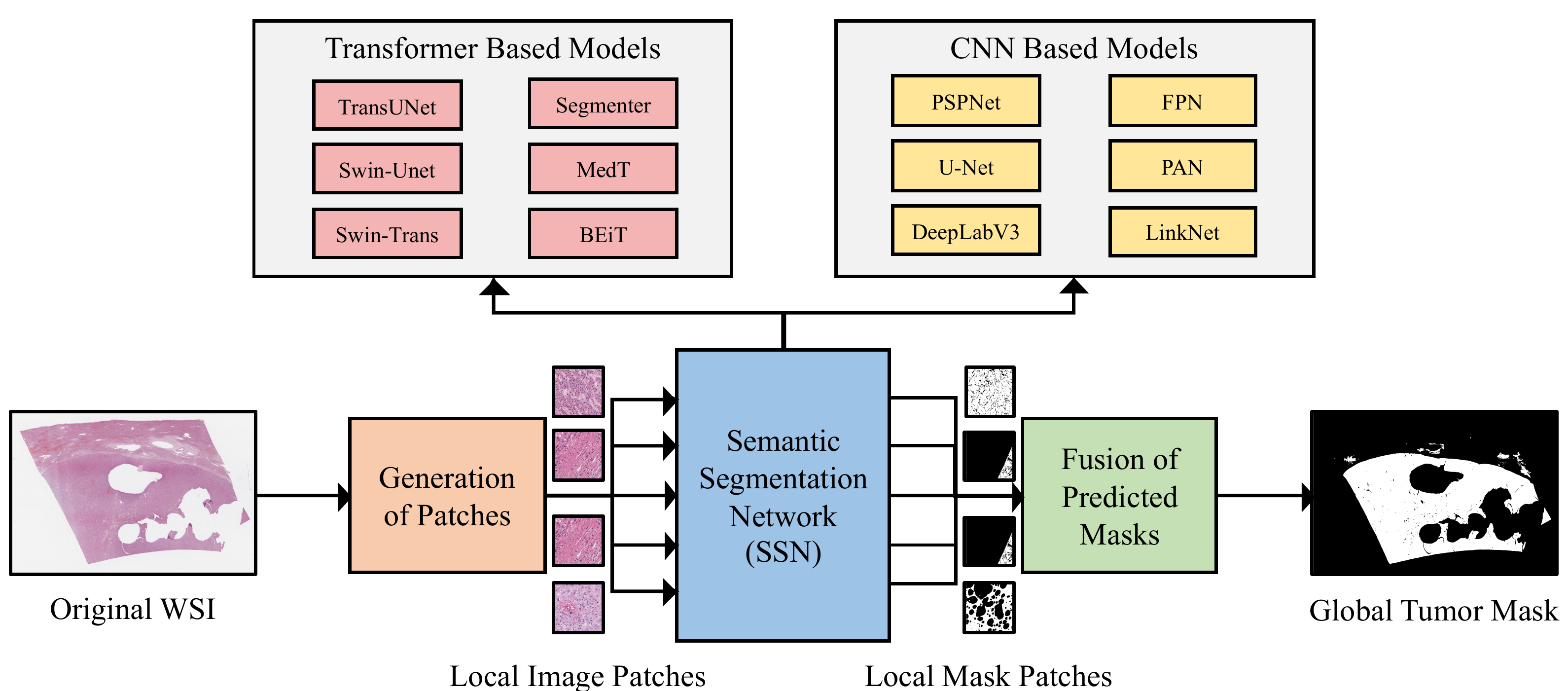}
    \caption{A pipeline of both CNN-based and Transformer-based cancer diagnosis}
    \label{fig:pipeline}
\end{figure}

Fig.\ref{fig:pipeline} shows the overall semantic segmentation pipeline for different approaches. To perform the semantic segmentation networks (SSN) on a Gigapixel histopathological image, the entire WSI is tiled into smaller patches at either 20$\times$ or 40$\times$ with the same patch size 512$\times$512. During the training stage, the patch sampler randomly crops training patches from the PAIP liver tumor dataset\cite{paip}. Since a WSI usually has large non-tissue areas that can be excluded from the computation, we discard patches in which more than half of the areas are background (i.e those with RGB values between 240-255). 

During the testing stage, to enable more smoothed global segmentation results, we allow for an overlap of neighboring patches, which is set to half of the patch size in both the horizontal and vertical directions. Each image patch is then fed into an SSN in order to achieve a predicted tumor segmentation mask. All image patches that belong to the same WSI are fused to a global segmentation mask with the same size as the original WSI. The performance of an SSN algorithm is evaluated by comparing the final global segmentation mask and the provided ground truth mask. For this purpose, we use the Jaccard index (also known as the IOU\cite{jaccard}), which is the ratio of the intersection of the predicted and ground truth masks to their union.

% \subsection{Training of SSN}
% \label{sec:Training of SSN}

% \begin{figure}[H]
%     \centering
%     \includegraphics[width=4.3in]{fig/Histopathology Report (9).pdf}
%     \caption{The algorithm flow for SSN training}
%     \label{fig:trainingSSN}
% \end{figure}

%  To teach SSN to predict tumors we supply it with cropped examples from of histopathological images and their groundtruth segmentation masks. These groundtruth masks are binary images that represent the true cancerous regions that have been carefully labelled by experienced pathologists. The cropped example images and their groundtruth masks that are fed to the network during training should have the same size as patches that are generated in Section \ref{sec:Cancer Segmentation Pipeline} in the prediction mode. The patch sampler randomly crops such exemplars from a training set of WSIs. Since a WSI usually has many blank areas that do not contain useful information we avoid cropping mostly white patches. Specifically, we select a patch only if the fraction of white pixels, i.e those with RGB values between 240-255, exceeds a given threshold. Furthermore, in each iteration of training, the network typically learns from not just one example but rather the whole batch of examples. In this work, we use batches that consist of 8 patches each.\\
 
% \subsection{Performance Metrics}
% \label{sec:Performance Metrics}

\subsection{Transformer based Methods}
\label{sec:Evaluated Models}

Six transformer-based SSN models, whose open-source code is available, are employed in this study:

\begin{enumerate}
    \item \textbf{TransUNet}\cite{tunet}: This model combines a Transformer with a CNN while still using the U-Net architecture. Specifically, they proposed a CNN-Transformer Hybrid Encoder. The transformers use the feature map, created by the CNN, as an input instead of a raw image. They also used a cascaded up-sampler as their decoder. In our experiment, we used R50-ViT-B\_16 which is a hybrid of ResNet-50 and ViT\cite{vit}.
    \item \textbf{Swin-Unet}\cite{sunet}: Unlike TransUNet, Swin-Unet is a purely transformer-based model with the encoder decoder architecture of U-Net. Like U-Net, this model has skip connections for local features. We used the initial checkpoint provided.  
    \item \textbf{Swin-Transformer}\cite{swin-trans}: This model is a hierarchical Transformer that can serve as a general backbone for various computer vision tasks. The representation is computed with shifted windows limiting self-attention computation to non-overlapping local windows. This makes the model efficient and flexible, also making it usable on image classification, object detection, and semantic segmentation. This is also a pure transformer. The open source code provides three configurations: tiny, small, and base. We report here the results of the tiny model. 
    \item \textbf{Segmenter}\cite{segm}: This is a fully transformer-based model. The encoder is made up of Multi-head Self Attention (MSA) blocks and Multi-Layer Perceptron blocks, as well as two layer norms before each block and residual connections after each block. Segmenter uses a linear decoder bilinearlly up-sampling the sequence into a 2D segmentation mask. The source code provides three models: small, base, and large. We report here the results of the base model. 
    \item \textbf{Medical Transformer}\cite{medt}: This model builds upon a basic Transformer-Based architecture with gated axial-attention blocks. The axial-attention blocks are composed of two self-attention blocks: one focusing on the width and the other focusing on the height. They are able to control the influence of the learned positional encodings on encoding non-local context.
    \item \textbf{BEiT}\cite{beit}: This method uses a transformer to learn a representation model through self-supervision training on ImageNet. They then use this representation as a backbone encoder and incorporate several deconvolution layers as a decoder to produce segmentation. In this experiment, we used the BEiT base model that was pretrained on ImageNet22K.
\end{enumerate}

\noindent To train these networks we used their provided open-source codes and fed them with data from our patch sampler. In each round of training, we fed 128 random examples to the network. These examples were randomly cropped from WSIs at a given level (40 magnification or 20 magnification) and put into batches of 8. Each model was trained for 1000 rounds. After each round, we computed the average loss of the model on the validation set. The model with the lowest loss value was retained and evaluated on the full set of the test slides using the Jaccard index. To eliminate the sensitivity to the learning rate, we trained each aforementioned model with different initial learning rates (0.01, 0.001, 0.0001) and only the best result is reported based on the validation set. 

\subsection{CNN based methods}
\label{sec:CNN based methods}
We also compared the above algorithms with major CNN-based semantic segmentation models including PSPNet\cite{psp}, U-Net\cite{unet}, DeepLabV3\cite{DeepLab}, FPN\cite{fpn}, PAN\cite{pan}, and LinkNet\cite{linknet}. These models were implemented using the open-source Semantic Segmentation Models package\cite{yakubovskiy}. 

% \section{Experiments}
% \label{sec:Experiments}

\section{Dataset and Experiments}
\label{sec:Dataset}

We employed the PAIP Grand Challenge dataset \cite{paip} to evaluate our histopathological image analysis algorithms. In this challenge, the participants had access to 100 WSIs of the liver collected by Seoul National University Hospital. Only 50 of them had ground truth segmentation masks publicly available as the other 50 were used for validation. Each image contains only a single tumor lesion from each patient. PAIP provides two types of tumor masks: viable and whole. The viable tumor is defined as the region containing the cell nest and the active cancerous parts of the tumor while the whole tumor contains the cell nests as well as the dead tissue surrounding the viable tumor and the tumor capsules. The results in this paper were obtained for the viable tumor type only. The slides were H\&E dyed, scanned by Aperio AT2 at $\times$20 power, and compressed into SVS format with the average size of 80,000$\times$80,000 pixels. We only used the 50 slides with ground truth in this work. Furthermore, we used 40 of those slides for training the SSN models while the remaining 10 were used for the testing. We also used a small validation set for selecting the best model during training. It consists of 150 patches randomly cropped from the test slides.\\

\begin{figure}[h]
    \centering
     \includegraphics[width=5in]{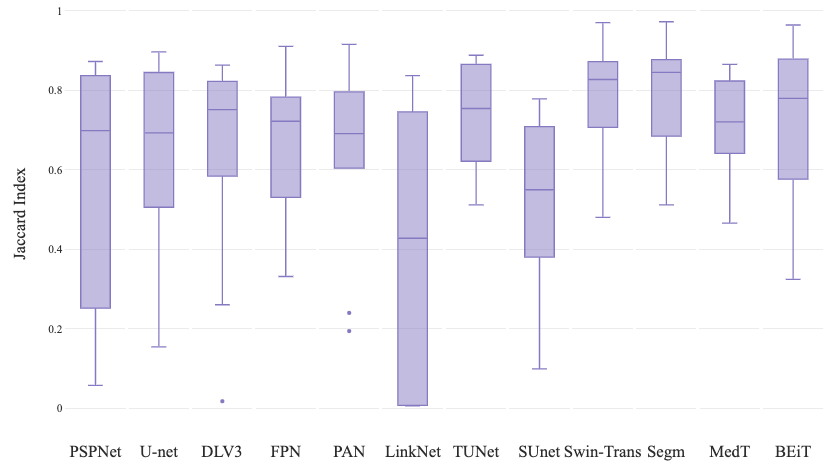}
    \caption{Tumor segmentation performance of different SSNs for patch size 512 and 40$\times$ magnification}
    \label{fig:ioubarchart}
\end{figure}

\section{Results}
\label{sec:Results}

\begin{figure}[H]
    \centering
    \includegraphics[width=6.9in]{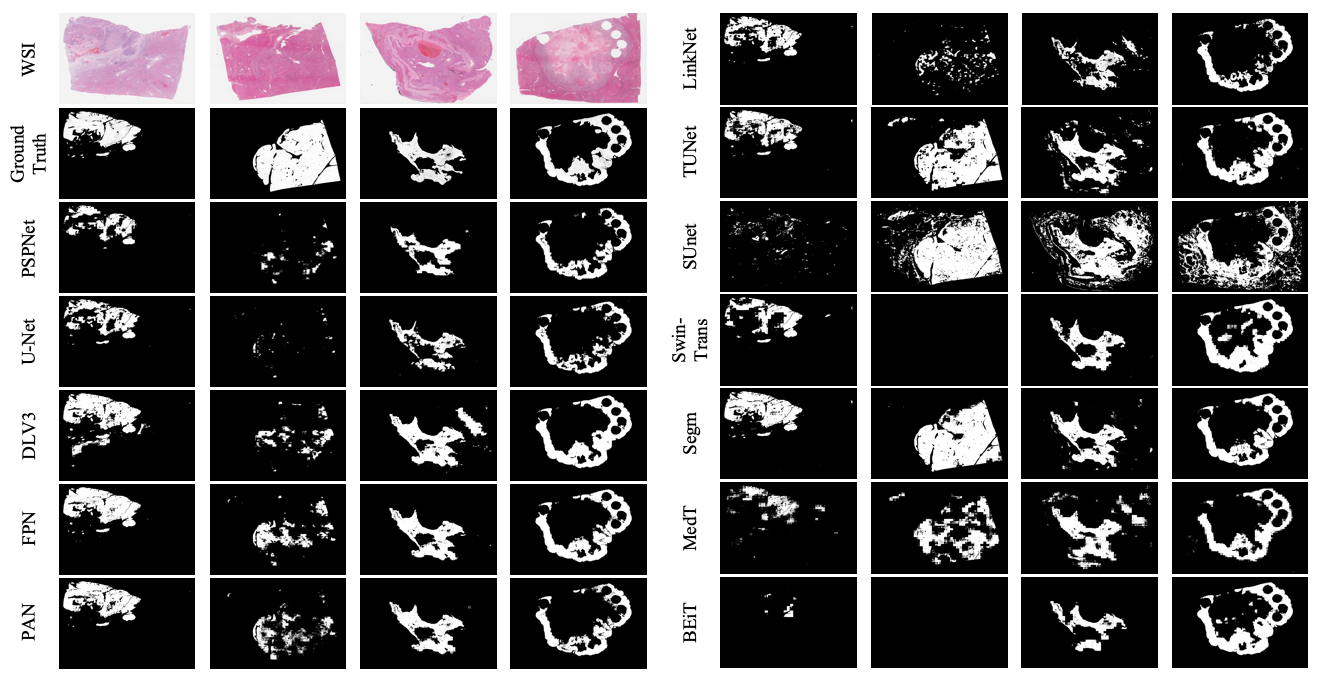}
    \caption{This figure presents the qualitative segmentation results using different approaches.}
    \label{fig:viableresultimages}
\end{figure}

The results are shown for the selected models in Table \ref{tab:iou} for the segmentation of viable tumors at the pyramid levels: 40$\times$ magnification and 20$\times$ magnification. The reported Jaccard index is averaged over all ten test slides. The chart in Fig.\ref{fig:ioubarchart} shows the average Jaccard index for different pyramid levels. Fig.\ref{fig:viableresultimages} shows the predicted tumor mask for some test slides along with their ground truth.

\begin{table}
    \centering
    \begin{tabular}{lcc}
    \hline
    CNN-based Models      & 40$\times$ Magnification   & 20$\times$ Magnification\\ \hline
     PSPNet\cite{psp} & 0.58 $\pm$ 0.33 & 0.49 $\pm$ 0.27\\ 
     U-Net\cite{unet} & 0.65 $\pm$ 0.24 & 0.60 $\pm$ 0.31\\
     DeepLabV3\cite{DeepLab} & 0.63 $\pm$ 0.28 & 0.67 $\pm$ 0.24\\
     FPN\cite{fpn} & 0.64 $\pm$ 0.20 & 0.72 $\pm$ 0.22\\
     PAN\cite{pan} & 0.63 $\pm$ 0.24 & 0.69 $\pm$ 0.23\\
     LinkNet\cite{linknet} & 0.35 $\pm$ 0.33 & 0.54 $\pm$ 0.25\\
     \hline
     Tranformer-Based Models  & 40$\times$ Magnification   & 20$\times$ Magnification\\ \hline 
     TransUNet(R50-ViT-B\_16)\cite{tunet} & 0.77 $\pm$ 0.12 & 0.77 $\pm$ 0.13\\
     Swin-Unet\cite{sunet} & 0.53 $\pm$ 0.23 & 0.42 $\pm$ 0.23\\
     Swin-Transformer(base)\cite{swin-trans} & 0.79 $\pm$ 0.14 & 0.71 $\pm$ 0.26\\
     Segmenter\cite{segm} & \textbf{0.80} $\pm$ 0.14 & \textbf{0.82} $\pm$ 0.11\\
     Medical Transformer\cite{medt} & 0.71 $\pm$ 0.14 & 0.62 $\pm$ 0.17\\
     BEiT\cite{beit} & 0.72 $\pm$ 0.21 & 0.66 $\pm$ 0.28\\ \hline
     \end{tabular}
    \caption{Average Jaccard index for viable tumors with a patch size of 512}
    \label{tab:iou}
\end{table}   
    
% \begin{table}[H]
%     \centering
%     \begin{tabular}{ccc}
%     \hline
%     \begin{flushleft}CNN-based Models\end{flushleft}      & 40$\times$ Magnification   & 20$\times$ Magnification\\ \hline
%      \begin{flushleft}PSPNet\cite{psp}\end{flushleft}    & 0.58        & 0.49\\ 
%      \begin{flushleft}U-Net \cite{unet}\end{flushleft}               & 0.65        & 0.60\\
%      \begin{flushleft}DeepLabV3 \cite{DeepLab}\end{flushleft}               & 0.63        & 0.67\\
%      \begin{flushleft}FPN\cite{fpn}\end{flushleft}                 & 0.64        & 0.72\\
%      \begin{flushleft}PAN\cite{pan}\end{flushleft}                & 0.63        & 0.69\\
%      \begin{flushleft}LinkNet  \cite{linknet}\end{flushleft}           & 0.35        & 0.54\\
%      \hline
%      \begin{flushleft}Tranformer-Based Models\end{flushleft} &  & \\
%      \hline
%      \begin{flushleft}TransUnet(R50-ViT-B\_16)  \cite{tunet}\end{flushleft}         & 0.77        & 0.77\\
%      \begin{flushleft}Swin-Unet\cite{sunet}\end{flushleft}           & 0.53        & 0.42\\
%      \begin{flushleft}Swin-Transformer(tiny)\cite{swin-trans}\end{flushleft}    & 0.79        & 0.71\\
%      \begin{flushleft}Segmenter(base)\cite{segm}\end{flushleft}           & 0.80        & 0.82\\
%      \begin{flushleft}Medical Transformer\cite{medt}\end{flushleft} & 0.71        & 0.62\\
%      \begin{flushleft}BEiT(base) \cite{beit}\end{flushleft}               & 0.72        & 0.66\\ \hline
%     \end{tabular}
%     \caption{Average Jaccard index for viable tumors with a patch size of 512}
%     \label{tab:iou}
% \end{table}

Several remarks can be made from the results:

\begin{enumerate}
    \item Transformer-based models tend to outperform FCN with the exception of Swin-Unet. This is consistent with published results in other application domains. 
    \item Our best average Jaccard index value 0.82 was obtained with Segmenter at patch size 512 magnification. Segmenter also achieved the highest scores at both resolution levels. 
    \item As shown by the results, Segmenter, Swin-Transformer, and TransUNet are the best performers. Both Segmenter and TransUNet are built upon the vision transformer VIT\cite{vit}, while Swin-Transformer is a hierarchical version of it. These three models were pretrained on ImageNet while other models such as Medical Transformer and BEiT were not. This could be a possible explanation for their superior performance. 
    \item While both Segmneter and Swin-Unet are fully transformer-based, Segmenter is the best model while Swin-Unet is the worst. Therefore, the performance most likely does not depend on whether the model is fully transformer-based or hybrid.
\end{enumerate}

\section{Discussion}
\label{sec:conclusion}
We have implemented an image analysis pipeline for tumor segmentation in whole-slide images. Using this pipeline we could evaluate the performance of different semantic segmentation models on a fair bases. Specifically, we quantitatively evaluated six state-of-the-art transformer-based models on the PAIP dataset and compared them with six major traditional CNN-based methods. The results showed that several transformer-based models indeed exhibit superior performance over the CNN-based models. In particular, Segmenter\cite{segm}, Swin-Transformer\cite{swin-trans} and TransUNet\cite{tunet}-all transformer-based-came out as the best performances among the twelve evaluated models.\\

\section{ACKNOWLEDGMENTS}       
This work has not been submitted for publication or presentation elsewhere.

% References
\bibliography{main} % bibliography data in report.bib
\bibliographystyle{spiebib} % makes bibtex use spiebib.bst

\end{document}